\documentclass[12pt]{article}
\usepackage[centertags]{amsmath}
\usepackage{amsfonts}
\usepackage{amssymb}
\usepackage{amsthm}
\usepackage{mathrsfs}

\newtheorem{thm}{Theorem}[section]

 \newtheorem{defn}[thm]{Definition}

\numberwithin{equation}{section}
\author {G. HACIBEK\.{I}RO\~{G}LU, M. \c{C}A\~{G}LAR \textrm{and} Y. POLATO\~{G}LU}
\title {The Higher Order Schwarzian Derivative: Its Applications for Chaotic Behavior and New Invariant Sufficient Condition of Chaos}    
\date{}

\begin{document}
\maketitle

\begin{abstract} The Schwarzian derivative of a function $f(x)$ which is defined in the interval $(a, b)$ having higher order derivatives is given by
$$\mathcal{S}_{f(x)}=\left( \frac{f''(x)}{f'(x)}\right)'- \frac{1}{2}\left( \frac{f''(x)}{f'(x)}\right)^2.$$
A sufficient condition for a function to behave chaotically is that its Schwarzian derivative is negative. In this paper, we try to find a sufficient condition for a non-linear dynamical system to behave chaotically. The solution function of this system is a higher degree polynomial. We define $n-$th Schwarzian derivative to examine its general properties. Our analysis shows that the sufficient condition for chaotic behavior of higher order polynomial is provided if its highest order three terms  satisfy an inequality which is invariant under the degree of the polynomial and the condition is represented by Hankel determinant of order $2$. Also the $n-$th order polynomial can be considered to be the partial sum of real variable analytic function. Let this analytic function be the solution of non-linear differential equation, then the sufficient condition for the chaotical behavior of this function is the Hankel determinant of order $2$ negative, where the elements of this determinant are the coefficient of the terms of $n, n-1, n-2$ in Taylor expansion. \end{abstract} 

\maketitle
\section{Introduction} We need following definition,
\begin{defn} If $f(x)$ is defined in the interval $(a, b)$ and had higher order derivatives on this interval, then the Schwarzian derivative of $f$ is defined by
\begin{equation}\nonumber\begin{split}\mathcal{S}_{f(x)} &= \left( \frac{f''(x)}{f'(x)}\right)'- \frac{1}{2}\left( \frac{f''(x)}{f'(x)}\right)^2 \\ &= \frac{f'''(x)}{f'(x)}- \frac{3}{2}\left( \frac{f''(x)}{f'(x)}\right)^2 \\ &= -2 \sqrt{f'(x)}\frac{d^2}{dx^2}\left( \frac{1}{\sqrt{f'(x)}} \right). \end{split}\end{equation}\end{defn}

Note that the most important properties of Schwarzian derivative are the following [1], [2], [6]:
\begin{enumerate}
\item The Schwarzian derivative is invariant under the bilinear transformation of $f(x)$, i.e, if
\begin{equation}\label{Eq1.0}T(f(x))=\frac{af(x)+b}{cf(x)+d}\end{equation}
then
$$\mathcal{S}_{T(f(x))}=\mathcal{S}_{f(x)}$$
and there are no transformations possessing this property. We also have 
$$\mathcal{S}_{g(t)}= \left( \frac{dx}{dt} \right)^2 \mathcal{S}_{f(x)} + \mathcal{S}_{\phi(t)},$$
where $x=\phi(t)$, $f(x)=f(\phi(t))=g(t)$.
\item The equation $\mathcal{S}_{f(x)}=0$ has the general solution $f(x)=\dfrac{ax+b}{cx+d}$ and the following hold:
$$\mathcal{S}_{f(x)} = -2 \sqrt{f'(x)}\frac{d^2}{dx^2}\left( \frac{1}{\sqrt{f'(x)}} \right) = 0 \Rightarrow \frac{d^2}{dx^2}\left( \frac{1}{\sqrt{f'(x)}} \right) = 0 \Rightarrow$$
$$f'(x)=\frac{1}{(ax+b)^2} \Rightarrow f(x)=\frac{c}{(ax+b)^2}.$$\end{enumerate}

Now we define the differential operator [3], [4], [5], [7]
\begin{equation}\label{Eq1.01}\begin{split}\mathcal{S}_{f_n(x)} &= \left( \frac{f^{(n+1)}(x)}{f^{(n)}(x)}\right)'- \frac{1}{(n+1)}\left( \frac{f^{(n+1)}(x)}{f^{(n)}(x)}\right)^2 \\ &= \left( \frac{f^{(n+2)}(x)}{f^{(n)}(x)} \right)- \frac{(n+2)}{(n+1)} \left( \frac{f^{(n+1)}(x)}{f^{(n)}(x)}\right)^2 \\ &= -(n+1) \sqrt[n+1]{f^{(n)}(x)} \frac{d^2}{dx^2}\left( \frac{1}{\sqrt{f^{(n)}(x)}} \right) \end{split}\end{equation}
which is called the $n-$th Schwarzian derivative.

The transformation $T_{f(x)}$ of $f(x)$ which leaves its $n-$th Schwarzian derivative invariant can be obtained in the following way: If $T_{f(x)}=f_1(x)$ then the relation
$$ -(n+1) \sqrt[n+1]{f^{(n)}(x)} \frac{d^2}{dx^2} \left( \frac{1} {\sqrt[n+1]{f^{(n)}(x)}} \right) = -(n+1) \sqrt[n+1]{f_1^{(n)}(x)} \frac{d^2}{dx^2} \left( \frac{1} {\sqrt[n+1]{f_1^{(n)}(x)}} \right) \Rightarrow$$
$$ \sqrt[n+1]{f^{(n)}(x)} \frac{d^2}{dx^2} \left( \frac{1} {\sqrt[n+1]{f^{(n)}(x)}} \right) =  \sqrt[n+1]{f_1^{(n)}(x)} \frac{d^2}{dx^2} \left( \frac{1} {\sqrt{f_1^{(n)}(x)}} \right),$$
having defined
\begin{equation}\label{Eq1.1} u(x)=\frac{1} {\sqrt[n+1]{f^{(n)}(x)}}, \quad v(x) = \frac{1} {\sqrt[n+1]{f_1^{(n)}(x)}}
\end{equation}
yields
\begin{equation}\label{Eq1.2} \frac{u''(x)}{u(x)}=\frac{v''(x)}{v(x)} \ \ \text{or} \ \ v''(x)- \frac{u''(x)}{u(x)}v(x)=0.
\end{equation}
So, the general solution of \eqref{Eq1.2} is given by 
\begin{equation}\label{Eq1.3}v(x)=u(x) \left( a_1 + a_2 \int_0^x \frac{dt}{(u(t))^2} \right),\end{equation}
where $a_1$ and $a_2$ are constants. On the other hand from \eqref{Eq1.1} and \eqref{Eq1.3} we have
$$f_1^{(n)}(x)=\frac{f^{(n)}(x)}{\left( a_1 + a_2 \int_0^x (f^{(n)}(x))^{\frac{2}{n+1}} dx \right)^{n+1}},$$
and hence
\begin{equation}\label{Eq1.4} f_1(x) = T(f(x)) =\underbrace{\int \int \int \cdots \int}_{n-\text{times}} \frac{f^{(n)}(x)}{\left( a_1 + a_2 \int_0^x (f^{(n)}(x))^{\frac{2}{n+1}} dx \right)^{n+1}} \underbrace{dx dx dx \cdots dx}_{n-\text{times}}.\end{equation}
For $n=1$, \eqref{Eq1.4} becomes the bilinear transformation of equation \eqref{Eq1.0}.

The Hankel matrix $H$ of the integer sequence $A=\{a_1, a_2, a_3, \cdots\}$ is the infinite matrix [8], [9]
\begin{displaymath}
{H} =
\left( \begin{array}{ccccc}
a_1 & a_2 & a_3 & a_4 & \ldots \\
a_2 & a_3 & a_4 & a_5 & \ldots \\
a_3 & a_4 & a_5 & a_6 & \ldots \\
\vdots & \vdots & \vdots & \vdots & \ddots
\end{array} \right).\end{displaymath}
Therefore the Hankel matrix $H_n$ of order of $A$ is the upper left $(n\times n)$ sub-matrix of $H$, and $h_n$, the Hankel determinant of order $n$ of $A$ is the determinant of the corresponding Hankel matrix of order $n$, $h_n=det(H_n)$. For example the Hankel matrix of order $4$ of the Fibonacci sequence $1, 1, 2, 3, 5, \cdots$, is  
\begin{displaymath}
{H_4} =
\left( \begin{array}{cccc}
1 & 1 & 2 & 3 \\
1 & 2 & 3 & 5 \\
2 & 3 & 5 & 8 \\
3 & 5 & 8 & 13  
\end{array} \right),\end{displaymath}
with $4-$th order Hankel determinant $h_4=0$.
\section{Main Results}
In this section we will give a sufficient condition for the chaotic behavior of an $n-$th degree polynomial. Let be a polynomial of degree $n$ with real  coefficient where $a_1 \neq 0$ having roots $x_i$, $i=1,2,3, \cdots, n$
\begin{equation}\label{Eq2.1} P(x) = a_1 x^n + a_2 x^{n-1} + a_3 x^{n-2} + \cdots + a_{n-1} x + a_n.\end{equation}
\begin{thm}\label{Thm2.1} The Schwarzian derivative of the polynomial of \eqref{Eq2.1} is
\begin{equation}\begin{split} \label{Eq2.2} \mathcal{S}_{P(x)} & = \frac{(n+2)!(n+1)a_1 \left(\frac{(n+2)!}{2!}a_1x^2+ (n+1)! a_2x + n!a_3\right)}{(n+1)\left( \frac{(n+2)!}{2!} a_1x^2 + (n+1)!a_2x+n!a_3 \right)^2}  \\ & \ \quad - \frac{(n+2) \left((n+2)!a_1x+ (n+1)! a_2\right)^2}{(n+1)\left( \frac{(n+2)!}{2!} a_1x^2 + (n+1)!a_2x+n!a_3 \right)^2}.
\end{split}\end{equation}\end{thm}
\begin{proof} The proof \eqref{Eq2.2} we will use the induction principle. Using formula \eqref{Eq1.01}, we get for $n=1$, $P(x)=a_1x^3+a_2x^2+a_3x+a_4$, i.e,
$$\mathcal{S}_{P(x)}=\left( \frac{P^{(n+1)}(x)}{P^{(n)}(x)} \right)' - \frac{1}{2} \left( \frac{P^{(n+1)}(x)}{P^{(n)}(x)} \right)^2 \Rightarrow$$
$$\mathcal{S}_{P(x)}= \frac{3! 2 a_1\left( \frac{3!}{2!}a_1 x^2 + 2! a_2x + 1! a_3\right) - 3 \left( 3! a_1 x + 2! a_2 x \right)^2}{2 \left( \frac{3!}{2!} a_1 x^2 + 2! a_2 x + 1! a_3  \right)^2},$$
so, \eqref{Eq2.2} is true for $n=1$.

Suppose that this result is true for $n=k$. Then we have, $P(x)=a_1x^{k+2}+ a_2 x^{k+1}+a_3x^k + a_4 x^{k-1} + \cdots + a_{k+1}$,
$$\mathcal{S}_{P(x)}=\left( \frac{P^{(k+1)}(x)}{P^{(k)}(x)} \right)' - \frac{1}{k+1} \left( \frac{P^{(k+1)}(x)}{P^{(k)}(x)} \right)^2 \Rightarrow$$
\begin{equation}\begin{split}\nonumber \mathcal{S}_{P(x)} & = \frac{(k+2)!(k+1)a_1 \left(\frac{(k+2)!}{2!}a_1x^2+ (k+1)! a_2x + k!a_3\right)}{(k+1)\left( \frac{(k+2)!}{2!} a_1x^2 + (k+1)!a_2x+k!a_3 \right)^2}  \\ & \ \quad -  \frac{(k+2) \left((k+2)!a_1x+ (k+1)! a_2\right)^2}{(k+1)\left( \frac{(k+2)!}{2!} a_1x^2 + (k+1)!a_2x+k!a_3 \right)^2}.
\end{split}\end{equation}
From the induction hypothesis, we then have, $P(x) = a_1 x^{k+3} + a_2 x^{k+2} + a_3 x^{k+1} + \cdots + a_{k+1},$
$$\mathcal{S}_{P(x)}=\left( \frac{P^{(k+2)}(x)}{P^{(k+1)}(x)} \right)' - \frac{1}{k+2} \left( \frac{P^{(k+2)}(x)}{P^{(k+1)}(x)} \right)^2 \Rightarrow$$
\begin{equation}\begin{split}\nonumber \mathcal{S}_{P(x)} & = \frac{(k+3)!(k+2)a_1 \left(\frac{(k+3)!}{2!}a_1x^2+ (k+2)! a_2x + (k+1)!a_3\right)}{(k+2)\left( \frac{(k+3)!}{2!} a_1x^2 + (k+2)!a_2x+(k+1)!a_3 \right)^2}  \\ & \ \quad - \frac{(k+3) \left((k+3)!a_1x+ (k+2)! a_2\right)^2}{(k+2)\left( \frac{(k+3)!}{2!} a_1x^2 + (k+2)!a_2x+(k+1)!a_3 \right)^2}.\end{split}\end{equation}
which is the desired conclusion.\end{proof}
\begin{thm}\label{Thm2.2} A necessary condition for an $n-$th degree polynomial to behave chaotically is that 
$$a_2^2-\frac{2n+4}{2n+3}a_1a_3>0 \ \ \text{for all} \ \ n=1, 2, 3, \cdots.$$\end{thm}
\begin{proof} Using the result of the Theorem \ref{Thm2.1}, we get
\begin{equation}\nonumber\begin{split} \mathcal{S}_{P(x)} < 0 & \Rightarrow \\ & \left[\frac{(n+2)!(n+1)}{2!}  - (n+2)!(n+2)\right]a_1^2 x^2 \\ & + \left[ (n+1)!(n+2)!(n+1) - 2 (n+1)!(n+3)!(n+2) \right]a_1a_2x  \\ & + \left[ (n)!(n+2)!(n+1)a_1a_3 - ((n+1)!)^2 (n+2)a_2^2 \right]<0 \end{split}\end{equation}
or 
\begin{equation}\label{Eq2.3}\begin{split} & \left[ ((n+2)!)^2(n+2) - \frac{((n+2)!)^2(n+1)}{2!}\right]a_1^2 x^2  \\ & + \left[ 2 (n+1)!(n+2)!(n+2) - (n+1)!(n+2)!(n+1)\right]a_1a_2x^2  \\ &+ \left[ ((n+1)!)^2 n! (n+2) a_2^2 - (n)!(n+2)!(n+1)a_1a_3 \right]>0. \end{split}\end{equation}
If we take 
\begin{equation}\nonumber\begin{split}  A &= \left[ ((n+2)!)^2(n+2) - \frac{((n+2)!)^2(n+1)}{2!}\right]a_1^2, \\ B &= \left[ 2 (n+1)!(n+2)!(n+2) - (n+1)!(n+2)!(n+1)\right]a_1a_2, \\ C &= \left[ ((n+1)!)^2 n! (n+2) a_2^2 - (n)!(n+2)!(n+1)a_1a_3 \right],\end{split}\end{equation}
then the inequality \eqref{Eq2.3} can be written in the form 
\begin{equation}\label{Eq2.4} Ax^2+Bx+C>0.\end{equation}
In order condition \eqref{Eq2.4} to be satisfied, the discriminant of the polynomial should be negative and the coefficient of $x^2$ should be positive, i.e, 
\begin{equation}\label{Eq2.5} A= \left[ ((n+2)!)^2(n+2) - \frac{((n+2)!)^2(n+1)}{2!}\right]a_1^2 >0 \end{equation}
i.e,
\begin{equation}\nonumber B^2-4AC=\left[ (n+2)!(n+3) - 2 (n+2)!(n+3)a_2 - 2(n+2)!(n+3)a_1a_3 \right]<0. \end{equation}
If we write \eqref{Eq2.5} for $n=1, 2, 3, \cdots$ we obtain following inequalities
\begin{equation}\nonumber\begin{split} 
n=1, \quad & a_2^2-3a_1a_3>0 \Rightarrow a_2^2-3a_1a_3 > 0, \\
n=2, \quad & 7a_2^2-8a_1a_3>0 \Rightarrow a_2^2-\frac{8}{7}a_1a_3 > 0, \\
n=3, \quad & 9a_2^2-10a_1a_3>0 \Rightarrow a_2^2-\frac{10}{9}a_1a_3 > 0, \\
n=4, \quad & 11a_2^2-12a_1a_3>0 \Rightarrow a_2^2-\frac{12}{11}a_1a_3 > 0, \\
n=5, \quad & 13a_2^2-14a_1a_3>0 \Rightarrow a_2^2-\frac{14}{13}a_1a_3 > 0, \\
\vdots \quad\quad & \quad\quad\quad\quad\vdots \quad\quad\quad\quad\quad\quad\quad\quad \vdots
\end{split}\end{equation}
so, by induction, we have
$$a_2^2-\frac{2n+4}{2n+3}>0, \quad n \geq 2.$$\end{proof}

As a result, a necessary condition for the chaotic behavior of the polynomial $P(x)$, is that
$$a_2^2-\frac{2n+4}{2n+3}a_1a_3>0, \quad n \geq 2.$$
Furthermore, letting $n$ go to infinity, we have $a_2^2-a_1a_3>0$. This implies that a real analytic function whose Maclaurin series expansion is $\sum_{n=0}^{\infty} a_n x^n$ behaves chaotically whenever $a_2^2-a_1a_3>0$. If we use the Hankel determinant, this condition can be written in the form 
$$\left| 
\begin{array}{lcr} 
a_1 & a_2 \\
a_2 & a_3
\end{array} 
\right| < 0.$$

\footnotesize

\textsc{G\"{u}rsel. Hac\i bekiro\~{g}lu}\\
Department of Physics, \\
TC \.{I}stanbul K\"{u}lt\"{u}r University, 34156 \.{I}stanbul, Turkey \\
e-mail: g.hacibekiroglu@iku.edu.tr\\

\textsc{Mert \c{C}a\~{g}lar}\\
Department of Mathematics and Computer Science, \\
TC \.{I}stanbul K\"{u}lt\"{u}r University, 34156 \.{I}stanbul, Turkey \\
e-mail: m.caglar@iku.edu.tr\\

\textsc{Ya\c{s}ar Polato\~{g}lu}\\
Department of Mathematics and Computer Science,\\
 TC \.{I}stanbul K\"{u}lt\"{u}r University, 34156 \.{I}stanbul, Turkey \\
e-mail: y.polatoglu@iku.edu.tr
\end{document}